\documentclass[11pt,twoside]{article}
\usepackage{asp2010}

\resetcounters

\begin{document}

\title{Data engineering for archive evolution}
\author{Rob~Seaman}
\affil{National Optical Astronomy Observatory, Tucson, AZ}

\begin{abstract}
From the moment astronomical observations are made the resulting data products begin to grow stale. Even if perfect binary copies are preserved through repeated timely migration to more robust storage media, data standards evolve and new tools are created that require different kinds of data or metadata. The expectations of the astronomical community change even if the data do not. We discuss data engineering to mitigate the ensuing risks with examples from a recent project to refactor seven million archival images to new standards of nomenclature, metadata, format, and compression.
\end{abstract}

\section{Data Preservation}
The archival preservation of astronomical data multiplies its scientific value.~\citep[{p.~143}]{nwnh_2010}  Astronomer Martin Harwit goes further to argue that archives are fundamental to scientific progress itself:
{\small \begin{em}``...we will need increasingly to ensure the long-term stability of data. Otherwise steady progress will remain beyond reach. In view of apparently inevitable evolutionary trends in the computer industry, astronomers and astrophysicists will need to devise fault-free methods for frequently transcribing records to maintain ready access to decade- or century-old data, including detailed descriptions of how those data were collected and reduced.''\end{em}}~\citep[{p.~361}]{harwit_2013}

In addition to creating a perpetual responsibility for making repeated verbatim copies to new media, preservation means active curation of data formats and of the evolving metadata needed to describe archival data products:
{\small \begin{em}``...when we no longer master the language in which data were stored, we may need to repeat observations or accept some interpretation of them on faith.  Unless observatories then become more powerful at more affordable cost, the rate at which we will be able to increase our understanding of the Universe may lag the rate at which previously gathered data become obsolete.''\end{em}}~(ibid., p.~348)

We will discuss issues of data formats and compression, the evolution of metadata, and the need to attach identifiers that retain meaning across diverse instrumentation.

\section{NOAO Science Archive}
Archiving of digital imaging data at the National Optical Astronomy Observatory began in 1993 (on July 20, a few days before the start of observing semester B) with ``Save the Bits'' ({\small STB})~\citep{2000ASPC..216..133S} - a system for tape-based, and later {\small CD-ROM}, data capture.  The initial online data store was developed for the NOAO Survey Archive~\citep{2003ASPC..295..100S}, and the earliest days of the NOAO Science Archive \textit{per se} began with the Data Cache Initiative~\citep{2005ASPC..347..679S} in 2004B.  In round numbers, 5+ million {\small FITS} files were captured to eighteen thousand tapes in the eleven year period between semesters 1993B and 2004A.  This has accelerated to 10+ million {\small FITS} files captured to spinning disk in the ten years between 2004B and 2014A.\footnote{Needless to say the average size of each image has increased dramatically as well.}  This most recent decade has seen development of new NOAO Science Archive~\citep{thomas_2009} features, and of the NOAO High Performance Pipeline System ({\small NHPPS})~\citep{2007ASPC..376..265S} that processes data sets from major instruments and submits new data products back to the archive.

Since data were being acquired and processed daily, each of these major system transitions and many other smaller evolutionary steps came with the likelihood of changes to the observatory data and metadata standards deployed in the archive.  In the case of NOAO these data are raw and processed optical or infrared astronomical images or spectra from about three dozen cameras or instruments on a dozen telescopes located on our three mountaintops in Arizona or Chile.\footnote{Over the years a variety of ancillary data products have also been archived, and there have been data sets from instrumentation on NOAO partner telescopes at other locations. These will not be discussed here.}
\begin{table}[ht]
\centering
\begin{tabular}{l l r r c c l}
\multicolumn{2}{c}{Semester} & \multicolumn{2}{c}{Holdings} \\
Begin & End & Number & TBytes & Name & Comp. & Details \\ [0.5ex]
1993B & 2004A & 5.0M & 40 & \emph{none} & \emph{none} & KP \& CT data on tape \\
1997B & 2004A & 0.2M & 20 & \emph{none} & \emph{none} & Mosaic 8k on 4meters \\
2004B & 2009B & 3.7M & 19 & serial & gzip & raw\\
2010A & 2013A & 3.1M & 59 & serial & {\small FPACK} & \\
2013B & & 0.4M & 29 & serial & {\small FPACK} & + {\small ASCII} header files \\
2014A & & 0.4M & 31 & {\small DSID} & {\small FPACK} & \\
2012B & 2014A & 0.2M & 97 & var. & {\small FPACK} & raw DECam \\
2012B & 2014A & 0.6M & 89 & var. & {\small FPACK} & DECam comm. pipeline \\
2004B & 2014A & 2.5M & 114 & var. & var. & {\small NHPPS} + CP \\
2007B & 2009B & 0.3M & 6 & serial & gzip & {\small NHPPS} \\
2004B & 2014A & 1.2M & 4 & serial & {\small PNG} & {\small NHPPS} preview images\\
\end{tabular}
\label{table:nonlin}
\caption{NOAO Science Archive legacy data sets by observing semester.  The archive mingles data from telescopes on Kitt Peak in Arizona, including the Mayall and WIYN 4m-class telescopes, with data from telescopes on Cerro Tololo and Cerro Pachon in Chile, including the Blanco and SOAR 4m telescopes.}\end{table}

Table 1 categorizes NOAO Science Archive holdings by various categories and by year and observing semester.  The data volume column (TBytes) is as converted to FITS tile compression using {\small FPACK}.  A total of 3.8 million raw and pipeline-reduced files were converted from the original gzip files, recovering 16.3 TBytes per archive copy.  The archive is replicated over geographically isolated copies and this corresponds to about 52 TBytes total savings.  Another 3.1 million files have been renamed to the new {\small DSID} standard and 600,000 {\small FITS} foreign encapsulated files have been unpacked into 1.2 million PNG files.  Various edge cases were also handled as indicated.  Total holdings are (conservatively) 15 million {\small FITS} files + 1.2 million {\small PNG} files + 16.2 million {\small ASCII} header files.

\section{Astronomical Data}
One can separate the idea of a mass data store from an archive.  In that case the initial data capture should make only minimal modifications to the original data products.  This was a basic design choice of the tape-based {\small STB}, whose mission was indeed ``saving the bits''.  The workflow included 1)~verifying data quality standards~\citep{2009ASPC..411....9S}, 2)~reformatting as a common format, 3)~computing checksums~\citep{seaman_2002}, 4)~attaching serial numbers, and 5)~adding metadata to a catalog.\footnote{These steps and other value-added handling (not described here) may not occur in this order, for instance some instruments write non-FITS data files and the automated conversion has always been implemented as an initial data preparation step before data are presented to the archive border component.}

An observatory and any partners will typically operate multiple mass data stores at geographically diverse locations separate from the telescope.  This implies that a Data Transport System ({\small DTS}) must exist to replicate data between sites.  The NOAO {\small DTS}~\citep{2010SPIE.7737E..1TF} has evolved from a very different data transport system implementation.  Most institutions will regard data transport to mean \emph{verbatim} replication, but in the general case transport could mean new names, a format conversion or change in compression, the generation of new checksums or attaching new timestamps.  More to the point replication opens the door for holdings of multiple mass stores to differ either intentionally or unintentionally, thus revealing a new requirement for periodic cross-verification of holdings.

The astronomical community benefits from widespread adoption of the {\small FITS} format (Flexible Image Transport System)~\citep{2010A&A...524A..42P} for images and binary tables, and this is NOAO's data format standard as well.  Over long periods of time even such stable standards may evolve~\citep{mink_adassxxiv}, and a data center should be prepared every few decades to completely reformat their entire and ever-growing back holdings.  While it may seem natural to consider integrating this with the more frequent migration of data onto new media, care should be taken to decouple logistical risks.\footnote{The most likely way to lose data in an archive is for the archivists to delete or overwrite it.~\citep{baker_2001}}

\subsection{FITS Tile Compression}
Changing the compression algorithm is less impactful than the underlying data format since services and libraries may remain the same, but it will still require rewriting all the data files.  For example, tools layered on the {\small CFITSIO}~\footnote{http://heasarc.nasa.gov/fitsio/} library can read files compressed with {\small FPACK}~\footnote{http://heasarc.nasa.gov/fitsio/fpack/} and gzip as well as uncompressed {\small FITS} files.  Motivations for using {\small FITS} Tile-Compression~\citep{2007ASPC..376..483S} are extensive, including lossless~\citep{2009PASP..121..414P} and lossy~\citep{2010PASP..122.1065P} regimes, as well as an option for tile-compressing binary tables~\citep{pence_2013}.  The point being that this complexity is inherent in the data themselves and the compression logistics need to be rich enough to support that complexity over long periods of time.

\subsection{Ancillary Data Products}
Images and tabular data are well supported by {\small FITS}, but a scientific archive will often contain ancillary data products of many types.  One example for the NOAO Science Archive is pipeline preview images.  {\small FITS} Foreign Encapsulation~\citep{zarate_2007} was designed to support the incorporation of such data sets.  Broader anticipated support for this format never materialized in the community and NOAO Science Data Management (SDM) has decided that extracting the {\small PNG}-format preview and thumbnail images from these files is appropriate.  This will allow the deployment of new features in the archive Portal, but as a result six hundred thousand {\small FITS} files have turned into 1.2 million {\small PNG} files (plus 1.2 million matching {\small ASCII} header files, see \S 4.4 below).

\section{Science Metadata}
The evolving data storage media, file formats and compression technologies are fundamental \textendash{} changing these may have implications for everything about the archive.  Scarcely less vital, however, are questions of metadata.  Each telescope and instrument has peculiarities of usage and a diversity of operational parameters that will only partially overlap prior facilities.  For example, the newly commissioned {\small KOSMOS}~\citep{seaman_2014c} spectrograph benefits from reusing the same Data Handling System used previously with the optical Mosaic and the infrared {\small NEWFIRM} wide-field imagers, but each of these variations results in a somewhat different set of {\small FITS} header keywords.  These then flow through pipeline-processing with the possibility of creating additional differences.  As a result, tools to aid the migration of legacy data sets may be required to support special cases for each of dozens of instrument and telescope combinations.

This metadata Babel can be at least partially mitigated if each instrument has a matching keyword dictionary~\citep{seaman_2014b}, and all adhere to common observatory~\citep{shaw_2009} and community standards.  However, just as individual archival images grow stale with time, so does the documentation of the data and metadata standards that describe each generation of data.  While special handling may apply generally to science metadata issues, there are four areas that require particular attention.

\subsection{Nomenclature}
Strong arguments can be made for choosing archive data set names strictly via a serial number heuristic \textendash{} human readability is not one of them.  NOAO SDM is renaming all archive holdings to a new user-friendly data set identifier ({\small DSID}) file naming convention~\citep{stobie_2013}, for example {\small \ttfamily c4d\_131212\_192824\_fri.fits.fz}~\citep{seaman_2014}.  This is a dense expression of metadata, with the initial {\small \ttfamily c4d} token indicating a {\small DEC}am observation made at the Blanco 4-meter telescope on Cerro Tololo in Chile.  The {\small \ttfamily fits.fz} file extensions\footnote {Note that tile-compressed FITS remains FITS and a simple {\ttfamily fits} extension would be perfectly correct.  The additional {\ttfamily fz} extension is merely an additional aid to human readability.  This is unlike the {\ttfamily gz} file extension for a gzip-compressed file, which is a completely different type of file from the original.} indicate a tile-compressed {\small FITS MEF} file.  This is a raw flat-field image ({\small \ttfamily fri}) and has a {\small UTC} timestamp on the given date and time.


\subsection{Timekeeping}
Issues associated with timekeeping are pervasive.  In general they divide between the syntactic \textendash{} most obviously issues related to the {\small Y2K} bug~\citep{bunclark_1997} or {\small ISO-8601} string formatting~\citep{ISO:2000:IDE}\footnote{Even international standards evolve, and ISO 8601:2000 was itself revised during the run-up to the millennium timekeeping crisis, and has since been replaced by ISO 8601:2004.} \textendash{} or semantic as pertaining to the definition of timescales, or their possible redefinition~\citep{seaman_2011}.

As the independent variable in most astronomical observations and often of key scientific importance especially in the time domain, date and timestamps, durations and time's angular equivalents, play a role in many other issues of scientific metadata.  Revising a time-related keyword may also have logistical implications, for instance in constructing filenames as discussed in \S4.1.  One of the inputs to build the NOAO {\small DSID} identifier is the observatory standard {\small DTUTC} keyword, which is a post-exposure timestamp added during data transport from the instrument to the first archival copy.  There are many edge cases, however, and post-exposure may mean a {\small UNIX} file creation timestamp versus reading the clock on demand when the archive transaction is triggered.  For cases of retroactively processing large numbers of images as considered here, neither one of these can apply, and a reliable timestamp must be constructed ex post facto from some instrument-dependent mix of keywords that may well have evolved during the lifetime of the instrument.  It is not simply a matter of taking the value of the {\small DATE-OBS} keyword \textendash{} not only since this changed with {\small Y2K}, but also because different instruments and telescopes have different failure modes for their clocks.

\subsection{Checksums}
Any modification to archive data holdings creates a necessity for recomputing checksums and other hash codes, such as the widely used {\small MD5} message digest.  This applies to both checksums that are maintained separately from the data files, for example as ingested into archive databases or services such as iRODS\footnote{http://irods.org} \textendash{} but also to embedded checksums like the FITS Checksum~\citep{seaman_2002} keywords.  Stale checksums must be kept in mind throughout intervening processing and will usually be updated during the final steps of any remediation project.

\subsection{Metadata as Data}
Early versions of the NOAO Science Archive relied on separate {\small ASCII} text header files to ingest image header keyword metadata into the archive databases.  This was a choice of convenience since {\small STB} already provided the feature of generating these files, and this avoided having to build {\small FITS} support into the Java classes for archive ingest.  When Java {\small FITS} support was later added to the archive support for the {\small ASCII} header files was dropped.  A few more years passed and a strategic decision was made to enable support for ancillary non-{\small FITS} data products as described in \S3.2, in particular relating to retiring support for {\small FITS} Foreign Encapsulation.  At this point it became clear that the original idea of representing the file metadata separately from the format \textendash{} {\small FITS} or otherwise \textendash{} had been the correct one all along.  What had been missing was a layer of meta-metadata (or purely logistical non-science metadata) that was now added as textual comments at the top of the {\small ASCII} FITS headers:

\begin{itemize}
\item[] {\scriptsize
\begin{verbatim}
% head -6 ./archive/mtn/20131212/ct4m/2012B-0001/c4d_131212_192824_fri.hdr
#filename = c4d_131212_192824_fri.fits.fz
#reference = c4d_131212_192824_fri.fits.fz
#filetype = TILED_FITS
#filesize = 691839360 bytes
#file_md5 = b28e108583ea0d72dc1f4baf2465377d
#hdr_version = 1.0
\end{verbatim}
}
\end{itemize}

Whether we call information data, metadata or even meta-metadata, it is all just {\em data} in the end.  As holdings are reformatted and remediated throughout a lengthy archive lifespan, particular aspects of those individual data products will need to be recomputed or replaced.  As a result files or database records will become more meta or less meta for a time or for a purpose.

\section{Remediation Workflow}
Putting all of the above together one sees that a toolkit paradigm is part of the solution.  Any archive will inevitably evolve in response to both internal and external pressures resulting in a diversity of holdings, as outlined in table 1 for the case of NOAO.

\begin{figure}
\center
\includegraphics[width=0.9\textwidth]{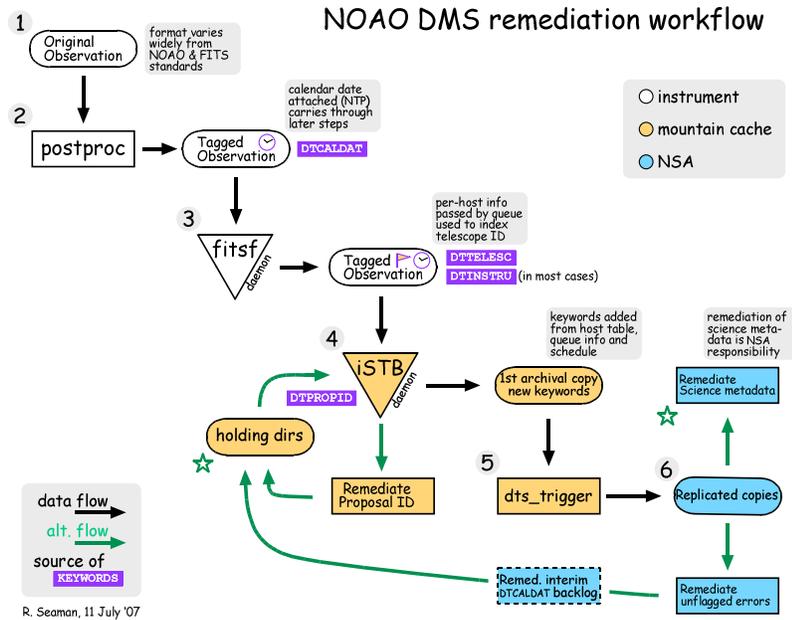}
\caption{Workflow for NOAO mountaintop remediation.}
\label{fig:trans}
\end{figure}

Figure 1 shows the initial steps involved in remediation of residual issues with non-standard or erroneous data or metadata.  Observations pass through several stages:
\begin{enumerate}
\item data handling system (DHS) creates original camera file
\item DHS triggers archive transaction, generates timestamp and datestamp (foreground)
\item FITS checksums and external knowledge (e.g., telescope) are logged (background)
\item data traverse network, first opportunity for remediation (human-mediated) \textendash{} missing observing proposal ID
\item second (automated) opportunity for remediation \textendash{} e.g., repair non-conforming FITS (faulty DATE-OBS or bad data record filler)
\item data are replicated to multiple downstream copies, ultimate responsibility for remediation is transferred (with data) to NOAO Science Archive
\end{enumerate}

The maintenance of this workflow is an exercise in progressively automating more of the steps \textendash{} for instance the yellow loop (steps 4 and 5) is a potential mountaintop embargo for data lacking an observing proposal identifier or a reliable observing calendar date.  Since the ID is fundamental to proprietary data authentication, it must be provided with a value in all cases.  The date is used to look up the proposal ID in the observing schedule\footnote{The data are also stored in a file system directory structure which includes subdirectories named by observing calendar date and the proposal ID.}

Several years ago the ``reliable-timekeeping'' project arranged to automatically attach the correct calendar date for data products generated by all instruments, that is, even if the instrument or telescope clock failed.  However, there are numerous use cases in which the proposal ID might not be reliably known (for example, multiple observers often split a night at the telescopes), and one-by-one these have been automated.  The alternative is for a human to make a judgment call after the fact.

However, as new instrumentation are commissioned (e.g., the Dark Energy Camera) issues like this must be addressed yet again and perhaps under completely new circumstances.  This process of becoming stale applies not only to the data, workflow and documentation, it even applies to the comic sans font used in this diagram.  The author was unable to update the font since the original vector drawing tool used to create the diagram is no longer being commercially supported.

Only very fundamental data and metadata issues can be addressed on the mountaintop at the same time as data-taking is occurring.  Needless to say that is also a never-ending discussion with the instrument teams and the observatory mountaintop operations staff to address the diverse causes of these issues.  All other remediation becomes the responsibly of the NOAO Science Archive along with other issues of long-term quality assurance.  

\subsection{Choosing the right tools}
These tools are deployed throughout the original data capture workflow from initial logging of transactions and checksums, through preprocessing of data sets\footnote{For instance using IRAF WFITS to reformat original data into the FITS format required by the archive.}, to {\small STB}, the archive boundary component.  Transformations of data that occur early in the workflow should be reversible, for example an {\small IRAF OIF} image that is reformatted as {\small FITS} can be recovered into its original format.  If metadata are updated it should only be by adding new keywords except under very carefully controlled circumstances.\footnote{Keyword values that are updated should be preserved by copying them into another keyword.}  Compression of raw data should generally (but perhaps not always) be lossless.  Long term preservation of data thus starts before the archive ever takes possession, and the earlier in the workflow an operation occurs the more robust the decision-making needs to be.  This will generally lead to relying on bedrock astronomical tools like {\small IRAF} or {\small CFITSIO}, or on widespread host tools like gzip or {\small MD5}.

Relying on victorinox\footnote{"Swiss Army Knife"} applications and standards has its own risks, however.  Gzip is a very poor compression format for astronomical data.  In addition to well known shortcomings of the {\small MD5} hash, it cannot be embedded in a FITS file.\footnote{Storing the MD5 for the corresponding image is one argument for a separate ASCII header file as described in \S4.4.  There remains no place to store the MD5 for the header file itself.}  A good rule of thumb would be to consider the strengths of the different tools \textendash{} and whether these are really what is needed.  For instance {\small MD5} is a cryptographic hash.  What parts of the data flow are subject to security concerns, versus what parts are subject to server and network risks to data integrity?  And gzip is a general-purpose compression tool optimized for text files, but astronomical data are predominantly binary integer and floating-point data with a gaussian-poisson noise model.  An analysis like this will quickly converge on design decisions similar to {\small FPACK} implementing {\small FITS} tile compression and the ones'-complement {\small FITS} checksum to provide efficient data representations across a large range of imaging and tabular astronomical data.

Not all data are well mapped to the {\small FITS} paradigm of scientific imaging and tabular data.  The choice is either to support each different flavor of ancillary data product separately, or rather to package these diverse data as {\small FITS} (or some other wrapper format).  Hence the need for tools implementing the {\small FITS} Foreign Encapsulation convention as discussed in \S3.2, namely the {\small IRAF FITSUTIL} package {\small FGWRITE} and {\small FGREAD} tasks.  As discussed above, policies for handling these may well change and a requirement to be able to read these files will persist indefinitely.  Separate requirements often interact, for instance NOAO has used encapsulated files to contain preview images that are automatically compressed by their {\small PNG} format.  The {\small FPACK} heuristics are smart enough to recognize header-data units that are something other than images or tables and skip over them, but the host gzip program squanders effort to compress these incompressible files.  The futility of gzipping  these data was recognized early on, but not until tens or hundreds of thousands of such doubly compressed files were created.  This then created the need to gunzip these files before using {\small FGREAD} to extract the original {\small PNG} files.

Whatever the nature of data holdings, archives are layered on relational databases to manage the corresponding metadata, typically via {\small SQL} commands.  Policy for the NOAO Science Archive is to keep the data synchronized with the metadata, thus updating a DB record implies the necessity to update a {\small FITS} header keyword(s).  Archives comprise numerous normalized DB tables, and indeed the NOAO Science Archive relies on more than one database, using an instance of i{\small RODS} to manage pathname mapping.  This results in two separate operations to load data into the archive, 1) register the data into i{\small RODS} using the ireg command, and 2) ingest it into the archive database using a command called {\small DSQUEUE}.\footnote{And indeed the archive itself comprises separate metadata and science databases, but this is an implementation detail.}  The fact of these multiple interacting databases means that the operation of updating the archive entry for a particular file is complex and subject to error, especially since each archive entry {\em per se} corresponds to two i{\small RODS} entries (one for the image and one for the {\small ASCII} header file).  And finally this implies that the most reliable way to update a complete archive entry is to first delete the old entry from the multiple databases (and possibly the old data from the disk), followed by reregistering and re-ingesting the updated image file.

Updating data holdings to match the evolving current standards is an arbitrary operation depending on the precise history of those holdings.  For instance we have discussed distinct sets of files whose names, formats, compression algorithm and header keywords require remediation in different combinations.  Tools were written and adapted to achieve each of these steps.  The {\small FPACK} (and corresponding {\small FUNPACK}) command for astronomical data compression has already been discussed, and similarly {\small FGREAD} to unencapsulate foreign (host) file formats.  A tool to update file names ({\small DSID}) was built on top of the {\small STB} routines that have been deployed to implement these new names on the mountaintops and for the pipelines.  A tool ({\small SBKEYS}) was needed to update the list of standard observatory keywords (a list which has grown over time) to the current header complement.  And finally a tool ({\small SBHDR}) creates the {\small ASCIII} header files when given a list of {\small FITS} files.\footnote{This also provides support for the {\small PNG} preview files when via the original encapsulated {\small FITS} files.}

\subsection{Using the tools right}
The broadest distinction between NOAO Science Archive holdings comes between 1)~raw data originating with mountaintop instrumentation, 2)~pipeline-processed data sets for major instruments, and 3)~investigator-processed data from the NOAO Survey Program\footnote{http://ast.noao.edu/activities/surveys}.  These survey data sets have to match validation requirements~\citep{seaman_2014a} whose own evolution will not be discussed here.

Each of these three classes of data are freighted with prerequisites before remediation and re-ingest are possible.  The most basic of these is that the observing proposal IDs for a particular data set must be first loaded into the NOAO Science Archive.  Many of the legacy data sets represent observing programs that have long been completed and the proposal metadata is likely itself to have become stale as investigators change institutions or email addresses, for instance.  This is a per-data-set, not per-image, operation but there are hundreds of NOAO proposals per semester and thus thousands of proposal IDs to wade through for any sizable backlog.

When prerequisites are satisfied, the workflow ties together the various tools in a sequence that will vary with the needs of the particular data set and contingent on data-driven issues (e.g., missing keywords or name collisions).  The real-time production workflow will be some sort of queue, the post-remediation workflow will be a batch-processing paradigm.  Thus tools for early remediation on the mountaintops will generally be scripts like {\small QUEUEPROC} that permit forcing a calendar date, proposal idea or queue name as appropriate.

While workflow tools for late remediation in the archive will be {\small IRAF} or unix scripts and compiled programs that are crafted as much as possible to operate in parallel.  Indeed, when converting millions of files from gzip to FITS tile compression format the only efficient possibility is to run dozens of instance simultaneously.  During such batch runs a premium is placed on monitoring of the numerous parallel batch jobs, and this is revealed in a need for additional tools like {\small SUM32} to verify FITS checksums.  Unfortunately, some steps are constrained to be executed on data files one at a time.  There is such a serial bottleneck on the final archive ingest operation, which is the ultimate throttle on how fast the archive can keep up with evolution.

\bibliography{P3-2.bib}

\end{document}